\documentclass[conference,oneside]{IEEEtran}

\usepackage{amsmath,amssymb,amsthm, graphicx,float,cite,bm}
\usepackage{algorithm,algorithmic,bbm}
\usepackage{bigints}

\allowdisplaybreaks[1]

\renewcommand{\d}[1]{\ensuremath{\operatorname{d}\!{#1}}}

\newtheorem{thm}{Theorem}
\newtheorem{defn}{Definition}
\newtheorem{rmk}{Remark}

\DeclareMathOperator{\tr}{tr}
\providecommand{\norm}[1]{\lVert#1\rVert}
\DeclareMathOperator{\diag}{diag}

\ifCLASSINFOpdf
\else
\fi
\begin{document}
\title{Controlled Sensing: A Myopic Fisher Information Sensor Selection Algorithm}

\author{\IEEEauthorblockN{Daphney-Stavroula Zois\thanks{This research has been funded in part by the following grants and organizations: ONR N00014-09-1-0700, NSF CNS-0832186,
CCF-0917343, CCF-1117896, CNS-1213128,  AFOSR FA9550-12-1-0215, DOT CA-26-7084-00, the National Center on Minority Health and Health Disparities (NCMHD) (supplement to P60 MD002254), Nokia and Qualcomm.} and Urbashi Mitra}
\IEEEauthorblockA{Ming Hsieh Dept. of Electrical Engineering\\
University of Southern California\\
Los Angeles, CA 90089-2565, USA\\
$\lbrace$zois, ubli$\rbrace$@usc.edu\\}}

% the acknowledgment of grants, issue a 
\IEEEoverridecommandlockouts


\maketitle

\begin{abstract}
%\boldmath
This paper considers the problem of state tracking with observation control for a particular class of dynamical systems. The system state evolution is described by a discrete--time, finite--state Markov chain, while the measurement process is characterized by a controlled multi--variate Gaussian observation model. The computational complexity of the optimal control strategy proposed in our prior work proves to be prohibitive. A suboptimal, lower complexity algorithm based on the Fisher information measure is proposed. Toward this end, the preceding measure is generalized to account for multi--valued discrete parameters and control inputs. A closed--form formula for our system model is also derived. Numerical simulations are provided for a physical activity tracking application showing the near--optimal performance of the proposed algorithm.
\end{abstract}

% \fi
\IEEEpeerreviewmaketitle

\section{Introduction}

In recent years, there has been an increasing interest in the problem of controlled sensing for inference in signal processing and related fields. In essence, the goal of controlled sensing is to characterize the way sensing modalities (\emph{e.g.} sensor type, number of samples) are used to accomplish a certain inference task. For instance, consider a physical activity tracking application in which the goal is to continuously estimate a person's physical activity (\emph{e.g.} walking, standing, \emph{etc}) using a set of sensors, such as accelerometers and heart--rate monitors. Intuitively, depending on which sensors are used, the
response may differ. Thus, carefully controlling the measurement process can dynamically refine the belief about the person's unknown time--evolving state and potentially lead to substantial performance gains. Applications of controlled sensing include, but are not limited to sensor management for object tracking \cite{AtiaTSP11}, ehealth \cite{ZoisTSP13}, spectrum sensing \cite{UnnikrishnanTSP10} and amplitude design for channel estimation \cite{RangarajanJSTSP07}.

We have previously considered the controlled sensing problem for inference in the case of discrete--time, finite--state Markov chains with controlled multi--variate Gaussian observations \cite{ZoisICASSP13}. In particular, we addressed the joint problem of deriving recursive formulae for a Minimum Mean--Squared Error (MMSE) state estimator and designing a control strategy to optimize its performance. In this regard, a Kalman--like estimator was designed and a dynamic programming (DP) algorithm optimizing the associated Mean--Squared Error (MSE) was derived.

It is usually the case that DP--based approaches suffer from the \emph{curse of dimensionality} (\emph{i.e.} one or more of the state, observation and control spaces are large) yielding no efficiently computable solutions. In our problem formulation, this fact is exacerbated by the fact that adopting the MSE as performance objective \cite{ZoisICASSP13} results in nonlinear cost functions.

Herein, as a first step toward the design of computationally efficient control strategies, we propose a sensor selection algorithm based on the Fisher information measure \cite{FisherPTRSL22}. This measure is extremely important in estimation theory and statistics since it 1) characterizes how well we can estimate a parameter based on a set of observations, and 2) is related to the concept of efficiency and the Kullback--Leibler divergence \cite{Gourieroux95}, which also constitutes a fundamental measure in information theory.

The problem of controlled sensing has been previously studied under time--invariant \cite{NitinawaratTAC13,NaghshvararAS13} and time--varying hypotheses \cite{KrishnamurthyTSP07, AtiaTSP11,ZoisTSP13}. In the latter case, it is common to assume that the unknown state is revealed through discrete noisy observations \cite{KrishnamurthyTSP07, ZoisTSP13}. On the other hand, the authors in \cite{AtiaTSP11} consider, among others, a Gaussian multi--variate signal model assuming i.i.d. measurements. In contrast, we consider time--varying systems with a generic Gaussian multi--variate signal model, which accounts for correlated measurements and enables fusion of multiple samples from different sensors.

In prior work, various performance objectives have been considered, \emph{e.g.} detection error probability and bounds \cite{ZoisTSP13,NitinawaratTAC13, NaghshvararAS13}, general convex distance measures \cite{KrishnamurthyTSP07, AtiaTSP11}, estimation bounds \cite{MasazadeCISS12}. In contrast, our focus is MSE, but since the optimal strategy is computationally intensive, we propose an algorithm based on the Fisher information measure. Various scalar functions of the Fisher information matrix have been previously considered as optimization criteria for sensor selection and active parameter/state estimation \cite{Scardovi05, HeroSJ11}. However, in all these cases, the differentiability of the associated likelihood function is implicitly assumed. In contrast, we consider \underline{multi--valued discrete parameters}, where this assumption fails.

Our contributions are as follows. To overcome the differentiability issue, we appropriately generalize the Fisher information measure. Furthermore, we derive a closed--form expression for our system model and propose a lower complexity algorithm that optimizes this expression with respect to control input selection. Finally, we evaluate the performance of the proposed algorithm using real data from a physical activity tracking application and show its near--optimal performance. 

\textbf{Notation.} Unless stated otherwise, all vectors are column vectors denoted by lowercase boldface symbols (\emph{e.g.} $\mathbf{x}$). On the other hand, matrices are denoted by uppercase boldface symbols (\emph{e.g.} $\mathbf{A}$). Sets are indicated by calligraphic symbols (\emph{e.g.} $\mathcal{X}$). $\tr(\cdot)$ denotes the trace operator, $\norm{\mathbf{x}}^{2}$ the $L^{2}$--norm of vector $\mathbf{x}$, $\mathbf{e}_{i}$ a vector of dimension determined by the context with $1$ in the $i$th position and zero everywhere else, $\diag(\mathbf{x})$ the diagonal matrix with elements the components of $\mathbf{x}$ and $|\mathbf{A}|$ the determinant of matrix $\mathbf{A}$.

\section{Problem Description}\label{sec:PD}

In this section, we describe the problem of controlled sensing. More precisely, we introduce our formulation, which includes our stochastic system model and the related stochastic optimization problem. For completeness, We also review our previously proposed Kalman--like system state estimator \cite{ZoisICASSP13}.

Consider a discrete--time dynamical system. Let $k = 0, 1, \dotsc$ denote discrete time and $\mathbf{x}_{k}$ denote a first--order Markov chain on the $n$--state state space $\mathcal{X}\doteq \lbrace \mathbf{e}_{1}, \mathbf{e}_{2}, \dotsc, \mathbf{e}_{n} \rbrace$ with $n$ indicating the total number of system states. We assume that the Markov chain dynamics are modeled by the transition probability matrix $\mathbf{P} = [P_{j|i}]_{n \times n}$ with $P_{j|i} = P(\mathbf{x}_{k+1} = \mathbf{e}_{j} | \mathbf{x}_{k} = \mathbf{e}_{i})$ and initial distribution $\bm{\pi} = [\pi_{i}]_{n \times 1}$ with $\pi_{i} = P(\mathbf{x}_{0} = \mathbf{e}_{i}), \forall \mathbf{e}_{i}, \mathbf{e}_{j} \in \mathcal{X}$. We also assume that the Markov chain is stationary, \emph{viz.} the associated transition probabilities do not change with time. We consider a set of sensors, which at each time step $k$, generate multiple noisy observations of the system state $\mathbf{x}_{k}$. A controller decides to receive all or any subset of these noisy observations by selecting an appropriate control input at time step $k-1$ denoted by $\mathbf{u}_{k-1}$. We assume that there is a finite number $\alpha$ of available controls, \emph{i.e.} $\mathbf{u}_{k-1} \in \mathcal{U} = \lbrace \mathbf{u}^{1}, \mathbf{u}^{2}, \dotsc, \mathbf{u}^{\alpha} \rbrace$. The resulting measurement vector is described by the following multivariate Gaussian model
\begin{equation}\label{eq:signal_model}
\mathbf{y}_{k} \big | \mathbf{e}_{i}, \mathbf{u}_{k-1} \sim f(\mathbf{y}_{k}|\mathbf{e}_{i},\mathbf{u}_{k-1}) = \mathcal{N}(\mathbf{m}_{i}^{\mathbf{u}_{k-1}}, \mathbf{Q}_{i}^{\mathbf{u}_{k-1}}),
\end{equation}

\noindent
where $\mathbf{m}_{i}^{\mathbf{u}_{k-1}}$ and $\mathbf{Q}_{i}^{\mathbf{u}_{k-1}}$ denote the mean vector and covariance matrix, respectively. To select a control input, the controller exploits the knowledge of the observation--control history $\mathcal{F}_{k} = \sigma \lbrace Y^{k}, U^{k-1} \rbrace$, where $\sigma \lbrace w \rbrace$ is the $\sigma$--algebra generated by $w$, $Y^{k} = \lbrace \mathbf{y}_{0}, \dotsc, \mathbf{y}_{k} \rbrace$ and $U^{k} = \lbrace \mathbf{u}_{0}, \dotsc, \mathbf{u}_{k} \rbrace$.

At each time step, the objective of the controller is to estimate the unknown system state by appropriately using the available observation and control input information. In that sense, the controller's operation can be divided into the following three phases: 1) \emph{control selection}, \emph{viz.} the appropriate $\mathbf{u}_{k}$ is selected based on $\mathcal{F}_{k}$, 2) \emph{measurement vector generation}, \emph{viz.} a measurement vector $\mathbf{y}_{k+1}$ is generated based on the selected control $\mathbf{u}_{k}$, and 3) \emph{system state estimation}, \emph{viz.} an estimate of the state $\mathbf{x}_{k}$ is determined based on the available information. In regard to the latter phase, we have recently proposed in \cite{ZoisICASSP13} an approximate MMSE system state estimator, which is formally similar to the well--known Kalman filter. Specifically, the posterior distribution of $\mathbf{x}_{k}$ given $\mathcal{F}_{k}$, denoted by $\mathbf{p}_{k|k} = [p_{k|k}^{1}, \dotsc, p_{k|k}^{n}]^{T}$ with $p_{k|k}^{i} = P(\mathbf{x}_{k} = \mathbf{e}_{i} | \mathcal{F}_{k})$, constitutes the optimal MMSE state estimate of the Markov chain system state. In  \cite{ZoisICASSP13}, we derived an approximate MMSE estimator $\hat{\mathbf{p}}_{k|k}$, which is characterized by a Kalman--like structure. Theorem \ref{thm:approx_MMSE_estimate} provides the associated equations.

\begin{thm}[Kalman--like estimator \cite{ZoisICASSP13}] \label{thm:approx_MMSE_estimate}
The Markov chain system state estimate at time step $k$ is recursively defined as
\begin{equation}\label{eq:filter_eqn}
\hat{\mathbf{p}}_{k|k} = \hat{\mathbf{p}}_{k|k-1} + \mathbf{G}_{k} [\mathbf{y}_{k} - \mathbf{y}_{k|k-1} ],~ k \geqslant 0
\end{equation}

\noindent
\begin{IEEEeqnarray}{rCl}
\text{with} \quad \quad \hat{\mathbf{p}}_{k|k-1} & = & \mathbf{P} \hat{\mathbf{p}}_{k-1|k-1},\\
\mathbf{y}_{k|k-1} & = & \mathcal{M}(\mathbf{u}_{k-1}) \hat{\mathbf{p}}_{k|k-1},\\
\mathbf{G}_{k} & = & \mathbf{\Sigma}_{k|k-1}\mathcal{M}^{T}(\mathbf{u}_{k-1}) \times (\mathcal{M}(\mathbf{u}_{k-1}) \times \nonumber \\ && \mathbf{\Sigma}_{k|k-1}\mathcal{M}^{T}(\mathbf{u}_{k-1}) + \widetilde{\mathbf{Q}}_{k})^{-1},
\end{IEEEeqnarray}

\noindent
where $\hat{\mathbf{p}}_{0|-1} = \bm{\pi}$, and $\bm{\pi}$ is the initial distribution over the system states, $\mathcal{M}(\mathbf{u}_{k-1}) = [\mathbf{m}_{1}^{\mathbf{u}_{k-1}},\dotsc,\mathbf{m}_{n}^{\mathbf{u}_{k-1}}]$, $\mathbf{\Sigma}_{k|k-1}$ is the conditional prediction error covariance matrix and $\widetilde{\mathbf{Q}}_{k} = \sum_{i=1}^{n} \hat{p}_{k|k-1}^{i} \mathbf{Q}_{i}^{\mathbf{u}_{k-1}}$.
\end{thm}

The estimator of Theorem \ref{thm:approx_MMSE_estimate} is employed during the system state estimation phase. It is evident from the related formulae that the estimation accuracy depends on the appropriate control input selection. Since the estimator's MSE performance is characterized by the \emph{conditional filtering error covariance matrix} $\mathbf{\Sigma}_{k|k} \doteq \mathbb{E} \lbrace (\mathbf{x}_{k} - \hat{\mathbf{p}}_{k|k}) (\mathbf{x}_{k} - \hat{\mathbf{p}}_{k|k})^{T} | \mathcal{F}_{k} \rbrace$, we formulated in \cite{ZoisICASSP13} the following optimization problem, which falls under the framework of partially observable Markov decision processes (POMDPs) \cite{BertsekasDPOC05}.

\vspace{5pt}
\noindent
\textbf{Controlled sensing problem.} Determine a sequence of control inputs $\mathbf{u}_{0}, \mathbf{u}_{1}, \dotsc, \mathbf{u}_{L-1}$, which minimizes the expected total cost
\begin{equation}\label{eq:total_cost}
J = \mathbb{E} \bigg \lbrace \sum_{k=1}^{L} \tr{\big (\mathbf{\Sigma}_{k|k}(\mathbf{y}_{k},\mathbf{u}_{k-1}) \big )} \bigg \rbrace,
\end{equation}

\noindent
where $L < \infty$.

\vspace{5pt}
The solution to the above problem is used during the control selection phase.

\section{Dynamic Programming Formulation}\label{sec:DPF}

In this section, we briefly summarize the solution of the controlled sensing problem proposed in our prior work \cite{ZoisICASSP13}. Specifically, having formulated our optimization problem as a POMDP enable us to seek a solution using dynamic programming principles, as illustrated in Theorem \ref{thm:DP_algorithm}.

\begin{thm}[\cite{ZoisICASSP13}] \label{thm:DP_algorithm}
For $k = L-1, \dotsc, 1,$ the \textit{cost--to--go} function $\overline{J}_{k}(\hat{\mathbf{p}}_{k|k-1})$ is related to $\overline{J}_{k+1}(\hat{\mathbf{p}}_{k+1|k})$ through the recursion
\begin{equation}
\begin{split}
\label{eq:DP_ss_bs_00}
&\overline{J}_{k}(\hat{\mathbf{p}}_{k|k-1}) = \underset{\mathbf{u}_{k-1} \in \mathcal{U}}{\min} \bigg [ \hat{\mathbf{p}}_{k|k-1}^{T} \mathbf{h}(\hat{\mathbf{p}}_{k|k-1}, \mathbf{u}_{k-1}) + \\& \int \mathbf{1}_{n}^{T} \mathbf{r}(\mathbf{y},\mathbf{u}_{k-1}) \hat{\mathbf{p}}_{k|k-1} \overline{J}_{k+1}\bigg ( \frac{\mathbf{P} \mathbf{r}(\mathbf{y},\mathbf{u}_{k-1}) \hat{\mathbf{p}}_{k|k-1}}{\mathbf{1}_{n}^{T}\mathbf{r}(\mathbf{y},\mathbf{u}_{k-1}) \hat{\mathbf{p}}_{k|k-1}} \bigg ) d \mathbf{y} \bigg ],
\end{split}
\end{equation}

\noindent
where $\mathbf{h}(\hat{\mathbf{p}}_{k|k-1}, \mathbf{u}_{k-1})$ is a column vector with components $h(\mathbf{e}_{1},\hat{\mathbf{p}}_{k|k-1},\mathbf{u}_{k-1}), \dotsc, h(\mathbf{e}_{n},\hat{\mathbf{p}}_{k|k-1},\mathbf{u}_{k-1})$ with $h(\mathbf{e}_{i},\hat{\mathbf{p}}_{k|k-1},\mathbf{u}_{k-1}) = 1 - \tr{\big( \mathbf{G}_{k}^{T} \mathbf{G}_{k} \mathbf{Q}_{i}^{\mathbf{u}_{k-1}} \big)} -  \norm{\hat{\mathbf{p}}_{k|k-1} + \mathbf{G}_{k} (\mathbf{m}_{i}^{\mathbf{u}_{k-1}} - \mathbf{y}_{k|k-1})}^2$, $\mathbf{1}_{n}$ is a vector of $n$ ones and $\mathbf{r}(\mathbf{y}_{k},\mathbf{u}_{k-1}) = \diag(f(\mathbf{y}_{k}|\mathbf{e}_{1},\mathbf{u}_{k-1}), \dotsc,f(\mathbf{y}_{k}|\mathbf{e}_{n},\mathbf{u}_{k-1}))$ is the $n \times n$ matrix of measurement vector pdfs. The cost--to--go function for $k = L$ is given by
\begin{equation}
\label{eq:DP_ss_bs_01}
\overline{J}_{L}(\hat{\mathbf{p}}_{L|L-1}) = \underset{\mathbf{u}_{L-1} \in \mathcal{U}}{\min} \bigg [ \hat{\mathbf{p}}_{L|L-1}^{T} \mathbf{h}(\hat{\mathbf{p}}_{L|L-1},\mathbf{u}_{L-1}) \bigg ].
\end{equation}
\end{thm}

Due to the complexity of the expressions involved, it is impossible to determine an analytical solution to (\ref{eq:DP_ss_bs_00}) and (\ref{eq:DP_ss_bs_01}).  Alternatively, we can numerically get an approximate solution. However, there are certain practical issues to consider. In particular, $\hat{\mathbf{p}}_{k|k-1}$ is continuous--valued, which implies that we must quantize the associated predicted belief space to acquire a finite number of states. In addition, the non--linear nature of the cost vector $\mathbf{h}(\hat{\mathbf{p}}_{k|k-1}, \mathbf{u}_{k-1})$ along with the multi--dimensional integration required during the computation of (\ref{eq:DP_ss_bs_00}) complicates the derivation of the DP policy. The above difficulties motivate our efforts for identifying a lower complexity scheme, which constitutes the major contribution of this work.

\section{Fisher Information}\label{sec:FI}

In this section, we review the Fisher information measure \cite{FisherPTRSL22} and determine its exact form for our system model. We also comment on its structure and individual characteristics.

\subsection{Definition}

The Fisher information \cite{FisherPTRSL22} constitutes a well--known information measure, which tries to capture the amount of information that an observable random variable $z \in \mathbb{R}$ contains about an unknown parameter $\theta \in \mathbb{R}$. It is related to the concept of efficiency in estimation theory since it provides a lower bound for the variance of estimators of a parameter, known as the Cram\'er--Rao lower bound (CRLB) \cite{VanTrees07}. To formally define the Fisher information, we begin with the following definition.

% Score function definition.
\begin{defn}[Score function \cite{SchervishTS95}]\label{defn:SF}
Let $f(z|\theta)$ be the conditional pdf of $z$ given $\theta$, which is also the likelihood function for $\theta$. For the observation $z$ to be informative about $\theta$, the density must vary with $\theta$. If $f(z|\theta)$ is smooth and differentiable, this change is quantified by the partial derivative with respect to $\theta$ of the natural logarithm of the likelihood function, i.e.
\begin{equation}\label{eq:SF}
S(\theta) = \frac{\partial}{\partial \theta} \ln f(x|\theta),
\end{equation}
\noindent
which is called the \textbf{score function}.
\end{defn}

\noindent
Under suitable regularity conditions (\emph{i.e.} differentiation with respect to $\theta$ and integration with respect to $z$ can be interchanged), it can be shown that the first moment of the score is zero
\begin{equation}
\mathbb{E} \lbrace S(\theta) \rbrace = \bigintssss \frac{f'(z|\theta)}{f(z|\theta)} f(z|\theta) \d{z} = \frac{\partial}{\partial \theta} \bigg \lbrace \int f(z|\theta) \d{z}  \bigg \rbrace = 0.
\end{equation}

\noindent
Next, we give the formal definition of Fisher information.

\begin{defn}[Fisher information \cite{SchervishTS95}]\label{defn:FI}
The variance of the score function $S(\theta)$ is the expected \textbf{Fisher information} about $\theta$, i.e.
\begin{equation}\label{eq:FI}
\mathcal{I}(\theta) = \mathbb{E} \lbrace S^{2}(\theta) \rbrace = \mathbb{E} \bigg \lbrace \bigg( \frac{\partial}{\partial \theta} \ln f(x|\theta) \bigg)^{2}\bigg \rbrace,
\end{equation}
\noindent
where $0 \leqslant \mathcal{I}(\theta) < \infty$.
\end{defn}

\noindent
We underscore that since the expectation of the score is zero, the associated term has been dropped in Definition~\ref{defn:FI}. Furthermore, we observe that the Fisher information characterizes the relative rate at which the pdf changes with respect to the unknown parameter $\theta$. In other words, the greater the expectation of a change is at a given value, the easier is to distinguish this value from neighboring values, and hence, we can achieve better estimation performance.

\subsection{Discrete Fisher Information}

We consider the dynamical system model in Section \ref{sec:PD}, where the unknown system state $\mathbf{x}_{k}$ is observed through a noisy measurement vector $\mathbf{y}_{k}$ that is shaped by a control input $\mathbf{u}_{k-1}$. Since the system state $\mathbf{x}_{k}$ corresponds to a discrete--time, finite--state Markov chain with $n$ states, we adopt hereafter the scalar notation $x_{k}$, where now $\mathcal{X} \doteq \lbrace 1, 2, \dotsc, n \rbrace$. 

In our formulation, there are three key components: 1) the system state $x_{k}$, which corresponds to the unknown parameter of interest, 2) the measurement vector $\mathbf{y}_{k}$ that refers to the observed random variable, and 3) the control input $\mathbf{u}_{k-1}$. Therefore, we need to ensure that these components are taken into consideration during the derivation of the Fisher information measure. First, we observe that the discrete nature of the system state $x_{k}$ prevents the direction application of Definitions \ref{defn:SF} and \ref{defn:FI}. To overcome this issue, we define the following \emph{generalized score function}
\begin{equation}\label{eq:GSF}
S(x_{k},x_{k}+h_{k},\mathbf{u}_{k-1}) = \frac{1}{h_{k}} \ln \bigg ( \frac{f(\mathbf{y}_{k}|x_{k} + h_{k}, \mathbf{u}_{k-1})}{f(\mathbf{y}_{k}|x_{k}, \mathbf{u}_{k-1})} \bigg ),
\end{equation}

\noindent
where the dependence on $\mathbf{u}_{k-1}$ has been stated explicitly and $h_{k}$ denotes a ``test point". The role of the latter is to avoid the need for differentiability imposed by Definition \ref{defn:SF}, while capturing any changes of the parameter values and enabling the computation of a generalized Fisher information measure. For completeness, we recall that the density of a multivariate Gaussian random vector $\mathbf{w} = [w_{1}, \dotsc, w_{d}]^{T}$ is given by
\begin{equation}\label{eq:MGD}
f(\mathbf{w}) = \frac{(2\pi)^{-d/2}}{|\mathbf{\Sigma}|^{1/2}} \exp \bigg ( -\frac{1}{2} (\mathbf{w} - \bm{\mu})^{T} \mathbf{\Sigma}^{-1} (\mathbf{w} - \bm{\mu}) \bigg ),
\end{equation}

\noindent
where $\bm{\mu}$ is the mean vector and $\mathbf{\Sigma}$ is the covariance matrix.

To determine, the exact form of the generalized score function for our system model, we substitute (\ref{eq:MGD}) in (\ref{eq:GSF}) and after some manipulations, we get
\begin{align}\label{eq:GSF_POMDP}
&S(x_{k},x_{k}+h_{k},\mathbf{u}_{k-1}) = \frac{1}{h_{k}}  \bigg [ \ln \bigg ( \sqrt{\frac{|\mathbf{\Sigma}_{x_{k}}^{\mathbf{u}_{k-1}}|}{|\mathbf{\Sigma}_{x_{k}+ h_{k}}^{\mathbf{u}_{k-1}}|}} \bigg ) \nonumber \\&-\frac{1}{2} \bigg( \mathbf{y}_{k}^{T} \mathbf{A}_{x_{k},x_{k}+h_{k}}^{\mathbf{u}_{k-1}} \mathbf{y}_{k} - 2 \mathbf{y}_{k}^{T} \mathbf{b}_{x_{k},x_{k}+h_{k}}^{\mathbf{u}_{k-1}} + c_{x_{k},x_{k}+h_{k}}^{\mathbf{u}_{k-1}} \bigg) \bigg ],
\end{align}

\noindent
where
\begin{align}
\mathbf{A}_{x_{k},x_{k}+h_{k}}^{\mathbf{u}_{k-1}}& \doteq \mathbf{\Sigma}_{x_{k}+h_{k}}^{\mathbf{u}_{k-1},-1} - \mathbf{\Sigma}_{x_{k}}^{\mathbf{u}_{k-1},-1} \\
\mathbf{b}_{x_{k},x_{k}+h_{k}}^{\mathbf{u}_{k-1}}& \doteq \mathbf{\Sigma}_{x_{k}+h_{k}}^{\mathbf{u}_{k-1},-1} \mathbf{m}_{x_{k}+h_{k}}^{\mathbf{u}_{k-1}} - \mathbf{\Sigma}_{x_{k}}^{\mathbf{u}_{k-1},-1} \mathbf{m}_{x_{k}}^{\mathbf{u}_{k-1}} \\
c_{x_{k},x_{k}+h_{k}}^{\mathbf{u}_{k-1}}& \doteq \mathbf{m}_{x_{k}+h_{k}}^{\mathbf{u}_{k-1},T} \mathbf{\Sigma}_{x_{k}+h_{k}}^{\mathbf{u}_{k-1},-1} \mathbf{m}_{x_{k}+h_{k}}^{\mathbf{u}_{k-1}} \nonumber \\&- \mathbf{m}_{x_{k}}^{\mathbf{u}_{k-1},T} \mathbf{\Sigma}_{x_{k}}^{\mathbf{u}_{k-1},-1} \mathbf{m}_{x_{k}}^{\mathbf{u}_{k-1}}.
\end{align}

\noindent
The expected value of the generalized score function in (\ref{eq:GSF_POMDP}) has the following form
\begin{align}
&\mathbb{E} \lbrace S(x_{k},x_{k}+h_{k},\mathbf{u}_{k-1}) \rbrace = \frac{1}{h_{k}} \bigg [ \ln \bigg ( \sqrt{\frac{|\mathbf{\Sigma}_{x_{k}}^{\mathbf{u}_{k-1}}|}{|\mathbf{\Sigma}_{x_{k}+ h_{k}}^{\mathbf{u}_{k-1}}|}} \bigg ) \nonumber \\ &-\frac{1}{2} \tr \big(\mathbf{A}_{x_{k},x_{k}+h_{k}}^{\mathbf{u}_{k-1}} \mathbf{\Sigma}_{x_{k}}^{\mathbf{u}_{k-1}}\big)  -\frac{1}{2} \mathbf{m}_{x_{k}}^{\mathbf{u}_{k-1},T}\mathbf{A}_{x_{k},x_{k}+h_{k}}^{\mathbf{u}_{k-1}}\mathbf{m}_{x_{k}}^{\mathbf{u}_{k-1}} \nonumber \\ &+ \mathbf{m}_{x_{k}}^{\mathbf{u}_{k-1},T}  \mathbf{b}_{x_{k},x_{k}+h_{k}}^{\mathbf{u}_{k-1}} -\frac{1}{2} c_{x_{k},x_{k}+h_{k}}^{\mathbf{u}_{k-1}} \bigg ] \doteq \frac{1}{h_{k}}\mu_{x_{k},x_{k}+h_{k}}^{\mathbf{u}_{k-1}},
\end{align}

\noindent
where we have exploited the following property for $\mathbf{w} \sim \mathcal{N}(\bm{\mu},\mathbf{\Sigma})$ \cite{Petersen08}
\begin{equation}\label{eq:property_355}
\mathbb{E} \lbrace \mathbf{w}^{T} \mathbf{A} \mathbf{w} \rbrace = \tr(\mathbf{A} \mathbf{\Sigma}) + \bm{\mu}^{T} \mathbf{A} \bm{\mu}.
\end{equation}

At this point, we define the following \emph{generalized Fisher information} measure
\begin{align}\label{eq:GFI}
\mathcal{I}(x_{k},x_{k} + h_{k},\mathbf{u}_{k-1})& \doteq \mathbb{E} \bigg \lbrace \bigg[ S(x_{k},x_{k}+h_{k},\mathbf{u}_{k-1}) \nonumber \\ &- \frac{1}{h_{k}}\mu_{x_{k},x_{k}+h_{k}}^{\mathbf{u}_{k-1}} \bigg]^{2} \bigg \rbrace,
\end{align}

\noindent
where once more the dependence on $\mathbf{u}_{k-1}$ has been stated explicitly. To determine the exact form of this measure for our system model, we exploit (\ref{eq:property_355}) along with the following properties for $\mathbf{w} \sim \mathcal{N}(\bm{\mu},\mathbf{\Sigma})$ \cite{Petersen08}
\begin{align}
\mathbb{E} \lbrace (\mathbf{w}^{T} \mathbf{A} \mathbf{w})^{2} \rbrace& = \tr(\mathbf{A}\mathbf{\Sigma}(\mathbf{A} + \mathbf{A}^{T})\mathbf{\Sigma}) + \bm{\mu}^{T} (\mathbf{A} + \mathbf{A}^{T}) \mathbf{\Sigma} \nonumber \\ &\times (\mathbf{A} + \mathbf{A}^{T}) \bm{\mu} + (\tr(\mathbf{A\Sigma}) + \bm{\mu}^{T}\mathbf{A}\bm{\mu})^{2}, \label{eq:property_825_d} \\
\mathbb{E} \lbrace \mathbf{w}^{T} \mathbf{A} \mathbf{w} \mathbf{w}^{T} \mathbf{b} \rbrace& = (\bm{\mu}^{T}\mathbf{A} + (\mathbf{A}\bm{\mu})^{T})\mathbf{\Sigma b} + (\tr(\mathbf{\Sigma A}^{T}) \nonumber \\ &+ \bm{\mu}^{T} \mathbf{A} \bm{\mu}) \mathbf{b}^{T} \bm{\mu}. \label{eq:property_825_g}
\end{align}

\noindent
Note that the above expressions can be simplified more in our case, since $(\mathbf{A}_{x_{k},x_{k}+h_{k}}^{\mathbf{u}_{k-1}})^{T} = \mathbf{A}_{x_{k},x_{k}+h_{k}}^{\mathbf{u}_{k-1}}$. After some manipulations, Eq. (\ref{eq:GFI}) becomes
\begin{align}\label{eq:GFI_POMDP}
&\mathcal{I}(x_{k},x_{k} + h_{k},\mathbf{u}_{k-1}) = \frac{1}{2h_{k}^{2}} \tr \bigg(\big(\mathbf{A}_{x_{k},x_{k}+h_{k}}^{\mathbf{u}_{k-1}} \mathbf{\Sigma}_{x_{k}}^{\mathbf{u}_{k-1}}\big)^{2}\bigg) \nonumber \\
&+ \frac{1}{h_{k}^{2}} \mathbf{b}_{x_{k},x_{k}+h_{k}}^{\mathbf{u}_{k-1}, T} \mathbf{\Sigma}_{x_{k}}^{\mathbf{u}_{k-1}} \mathbf{b}_{x_{k},x_{k}+h_{k}}^{\mathbf{u}_{k-1}} -\frac{1}{h_{k}^{2}} \mathbf{m}_{x_{k}}^{\mathbf{u}_{k-1}, T} \mathbf{A}_{x_{k},x_{k}+h_{k}}^{\mathbf{u}_{k-1}} \nonumber \\
& \times \bigg [ -2\ln \bigg ( \sqrt{\frac{|\mathbf{\Sigma}_{x_{k}}^{\mathbf{u}_{k-1}}|}{|\mathbf{\Sigma}_{x_{k}+ h_{k}}^{\mathbf{u}_{k-1}}|}} \bigg ) \mathbf{m}_{x_{k}}^{\mathbf{u}_{k-1}} + \mathbf{\Sigma}_{x_{k}}^{\mathbf{u}_{k-1}} \mathbf{A}_{x_{k},x_{k}+h_{k}}^{\mathbf{u}_{k-1}} \mathbf{m}_{x_{k}}^{\mathbf{u}_{k-1}} \nonumber \\
& -2 \mathbf{\Sigma}_{x_{k}}^{\mathbf{u}_{k-1}} \mathbf{b}_{x_{k},x_{k}+h_{k}}^{\mathbf{u}_{k-1}} \bigg ].
\end{align}

\noindent
We notice that, as expected, the resulting measure constitutes a complicated function of the statistics of the underlying multivariate Gaussian model. At the same time, these statistics are driven by the selected control input $\mathbf{u}_{k-1}$.

As already discussed, the generalized Fisher information measure in (\ref{eq:GFI}) avoids the need for differentiability of the associated likelihood function by using test points. In general, these test points are selected so that the resulting parameter space is covered, yet ensuring that invalid parameter values are ignored. For our problem, this implies that test points should be selected to account for the discrete nature of the parameter space, \emph{i.e.} test points should be state--dependent: $h_{t} \in \mathcal{A} \doteq \big \lbrace h_{t}(x_{t}) \in \mathbb{R} ~|~ x_{t} + h_{t}(x_{t}) \in \mathcal{X} \big \rbrace$. For example, if we assume $\mathcal{X} = \lbrace 1,2,3,4 \rbrace$, the valid test point values for state $x_{k} = 2$ are $-1, 1$ and $2$.

\begin{rmk}
The generalized Fisher information measure in (\ref{eq:GFI}) can also be employed for constrained parameters. In that case, the selection of test points is less restrictive than our case, viz. if $\theta$ is defined in a constrained interval of the form $[a,b]$, $h_{k}$ is selected so that $\theta + h_{k} \in [a,b]$. 
\end{rmk}

\begin{algorithm}[tb!]
\caption{Greedy Fisher Information Sensor Selection (GFIS$^{2}$)}
\begin{algorithmic}[1]
\STATE // \textsc{Initialization}
\STATE $\hat{\mathbf{p}}_{0|-1} := \pi$, $\hat{x}_{0} = \arg \max \hat{\mathbf{p}}_{0|-1}$;
\STATE Determine $\phi(\hat{x}_{0},\mathbf{u}_{-1})$ using (\ref{eq:GFI_POMDP});
\STATE $\mathbf{u}_{-1}^{GFIS^2} = \arg \max \big[ \phi(\hat{x}_{0},\mathbf{u}_{-1}) \big]$;
\STATE // \textsc{Main Loop}
\FOR{$k=0:L$}
\STATE Request measurement vector $\mathbf{y}_{k}$ based on $\mathbf{u}_{k-1}$;
\STATE $\mathbf{y}_{k|k-1} = \mathcal{M}(\mathbf{u}_{k-1}) \hat{\mathbf{p}}_{k|k-1}$;
\STATE $\mathbf{G}_{k} = \mathbf{\Sigma}_{k|k-1}\mathcal{M}^{T}(\mathbf{u}_{k-1}) (\mathcal{M}(\mathbf{u}_{k-1})\mathbf{\Sigma}_{k|k-1}\mathcal{M}^{T}(\mathbf{u}_{k-1}) + \widetilde{\mathbf{Q}}_{k})^{-1}$;
\STATE $\hat{\mathbf{p}}_{k|k} = \hat{\mathbf{p}}_{k|k-1} + \mathbf{G}_{k}(\mathbf{y}_{k} - \mathbf{y}_{k|k-1})$;
\STATE // \textsc{System state estimate at time step} $k$
\STATE Declare system state as: $\hat{x}_{k} := \arg \max \hat{\mathbf{p}}_{k|k}$;
\STATE $\hat{\mathbf{p}}_{k+1|k} = \mathbf{P} \hat{\mathbf{p}}_{k|k}$; $\hat{x}_{k+1} = \arg \max \hat{\mathbf{p}}_{k+1|k}$;
\STATE Determine $\phi(\hat{x}_{k+1},\mathbf{u}_{k})$ using (\ref{eq:GFI_POMDP})
\STATE $\mathbf{u}_{k}^{GFIS^2} = \arg \max \big[ \phi(\hat{x}_{k+1},\mathbf{u}_{k}) \big]$;
\STATE $k := k + 1$;
\ENDFOR
\end{algorithmic}
\label{alg:GFIS2}
\end{algorithm}

\section{Greedy Fisher Information Sensor Selection}\label{sec:GFIS2}

In this section, we propose a myopic sensor selection algorithm that exploits the generalized Fisher information measure (\ref{eq:GFI_POMDP}) and discuss about its implementation.

As already discussed, Fisher information captures the amount of information that an observable random variable carries about an unknown parameter. Ideally, we would like to maximize this information so that we can infer with certainty the value of the unknown parameter of interest. In our formulation, we also have an extra degree of freedom, the control input, which we can exploit accordingly to maximize the amount of information we acquire with respect to the unknown parameter. To this end, we propose the following myopic sensor selection strategy that maximizes the generalized Fisher information measure (\ref{eq:GFI_POMDP}) at each time step
\begin{equation}\label{eq:GFIS2}
\mathbf{u}_{k}^{GFIS^2} = \arg \max \big[ \phi(x_{k+1},\mathbf{u}_{k}) \big],
\end{equation}

\noindent
where $\phi(x_{k+1},\mathbf{u}_{k}) \doteq \max_{h_{k+1}}[ \mathcal{I}(x_{k+1},x_{k+1} + h_{k+1},\mathbf{u}_{k}) ]$. We underscore that the Fisher information in (\ref{eq:GFI_POMDP}) is maximized with respect to all possible test point values at each time step to ensure that the tightest Fisher information is computed.

Examining Eq. (\ref{eq:GFIS2}), we notice that the function $\phi(\cdot)$ depends both on the system state $x_{k+1}$ and the control input $\mathbf{u}_{k}$. However, the former variable is unknown, in fact this is what wish to infer. To overcome this impediment, we instead use an estimate of the system state, \emph{i.e.} 
\begin{equation}
\hat{x}_{k+1} = \arg \max \hat{\mathbf{p}}_{k+1|k}, 
\end{equation}
 
\noindent
where $\hat{\mathbf{p}}_{k+1|k}$ is computed through our Kalman--like filter recursions of Theorem \ref{thm:approx_MMSE_estimate}. Our proposed strategy, which we refer to as Greedy Fisher Information Sensor Selection (GFIS$^{2}$), is shown in Algorithm \ref{alg:GFIS2}. Note that the sensor selection part is intertwined with the Kalman--like filter recursions. In particular, at each time step, GFIS$^{2}$ determines the predicted belief state, which then uses to determine the appropriate control input via (\ref{eq:GFIS2}).

The proposed algorithm presents several benefits. Among these, the most important is its myopic structure, \emph{i.e.} no computation of an expected future cost is any more required. As a result, the proposed algorithm incurs much lower computational complexity compared to DP. At the same time, the computation of the values of the function $\phi(\cdot)$ along with the optimization step (\ref{eq:GFIS2}) can be completed off--line. Consequently, the proposed strategy can be implemented as a look--up table, suggesting a very efficient implementation. The associated complexity is $\mathcal{O}(n \alpha c_{GFI})$, where $c_{GFI}$ is the complexity of computing $\phi(\cdot)$ for a pair $(x_{k+1},\mathbf{u}_{k})$, versus $\mathcal{O}((d+1)^{n} \alpha c_{\mathbf{h}, int})$ for DP, where $(d+ 1)^{n}$ is the number of predicted belief states\footnote{We quantize the predicted belief space with resolution $d$.} and  $c_{\mathbf{h}, int}$ the complexity of computing the term inside the minimization of (\ref{eq:DP_ss_bs_00}) for a pair $(\mathbf{p}_{k|k-1},\mathbf{u}_{k})$.

\begin{figure}[tb!]
\begin{center}
\noindent
  \includegraphics[width=\linewidth]{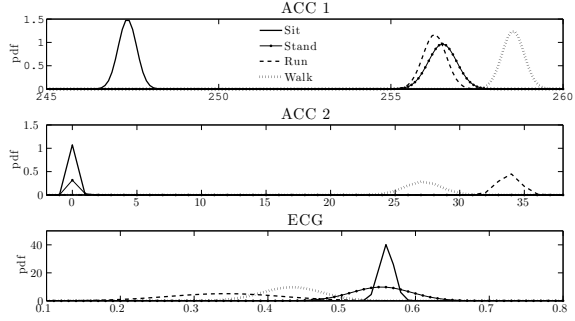}
  \caption{Distributions and four physical activity states and three sensors for a single participant.}\label{fig:4Hypothesis}
\end{center}
\end{figure}

\section{Numerical Results}

In this section, we provide numerical simulations to evaluate the GFIS$^2$ algorithm developed in Section \ref{sec:GFIS2}. We also compare its performance with the DP algorithm of Section \ref{sec:DPF}. Both algorithms are applied to a physical activity state tracking example also considered in \cite{ZoisICASSP13}.

We consider a Wireless Body Area network (WBAN) that consists of three heterogeneous sensors (two accelerometers (ACCs)\footnote{ACC 1 is an internal phone sensor, whereas ACC 2 is a standalone sensor and they are positioned in different parts of the body.}, one electrocardiograph (ECG)) deployed in a simple star topology and a fusion center, which is a Nokia N95 mobile phone. The sensors measure the vital signs of an individual, who is moving between four physical activity states: \emph{Sit}, \emph{Stand}, \emph{Run} and \emph{Walk}. The goal of the fusion center is to continuously estimate the individual's unknown physical activity state by adaptively communicating with the sensors in its WBAN. Thus, at each time step, it selects a control input $\mathbf{u}_{k-1} = [N_{1}^{\mathbf{u}_{k-1}},N_{2}^{\mathbf{u}_{k-1}},N_{3}^{\mathbf{u}_{k-1}}]$, where $N_{1}^{\mathbf{u}_{k-1}}, N_{2}^{\mathbf{u}_{k-1}}, N_{3}^{\mathbf{u}_{k-1}}$ indicate the number of samples requested from ACC 1, ACC 2, and ECG, respectively. We assume that $\sum_{i=1}^{3} N_{l}^{\mathbf{u}_{k-1}} \leqslant N$, where $N$ is fixed. The mean vectors and covariance matrices in (\ref{eq:signal_model}) have the form \cite{ZoisICASSP13}
\begin{align}
\mathbf{m}_{i}^{ \mathbf{u}_{k-1} }& = [\mu_{i,S_{1}}^{ \mathbf{u}_{k-1} }, \mu_{i,S_{2}}^{ \mathbf{u}_{k-1} }, \mu_{i,S_{3}}^{ \mathbf{u}_{k-1} }]^T, \\
\mathbf{Q}_{i}^{ \mathbf{u}_{k-1} }& = \diag (\mathbf{Q}_{i}^{ \mathbf{u}_{k-1} }(S_{1}), \mathbf{Q}_{i}^{ \mathbf{u}_{k-1} }(S_{2}), \mathbf{Q}_{i}^{ \mathbf{u}_{k-1} }(S_{3}))
\end{align}

\noindent
where $S_{\ell}$ denotes sensor $\ell$. For a particular sensor $S_{\ell}$, the mean vector $\mu_{i,S_{\ell}}^{ \mathbf{u}_{k-1} }$ is of size $N_{\ell}^{ \mathbf{u}_{k-1} } \times 1$ and the covariance matrix is defined as $\mathbf{Q}_{i}^{ \mathbf{u}_{k-1} }(S_{\ell}) = \frac{\sigma_{S_{\ell},i}^{2}}{1-\phi^2} \mathbf{T} + \sigma_{z}^{2} \mathbf{I}$, where $\mathbf{T}$ is a Toeplitz matrix with first row/column $[1,\phi,\phi^2,\dotsc,\phi^{N_{\ell}^{\mathbf{u}_{k-1}} -1}]$, $\mathbf{I}$ is the $N_{\ell}^{ \mathbf{u}_{k-1} } \times N_{\ell}^{ \mathbf{u}_{k-1} }$ identity matrix, $\phi$ is the model parameter and $\sigma_{z}^{2}$ accounts for sensing and communication noise.

Fig.~\ref{fig:4Hypothesis} shows the underlying distributions for the four physical activity states for a single participant for each of the three sensors. In the simulations, we have adopted $N = 2$ but are methods are directly applicable to larger values of $N$. The Markov chain transition probability matrix used is $\mathbf{P} = [0.6 ~0.2 ~0 ~0.4; 0.1 ~0.4 ~0.1 ~0; 0 ~0.1 ~0.3 ~0.3; ~0.3 ~0.3 ~0.6 ~0.3]$.

\begin{figure}[tb!]
\centering
{\includegraphics[width=\columnwidth]{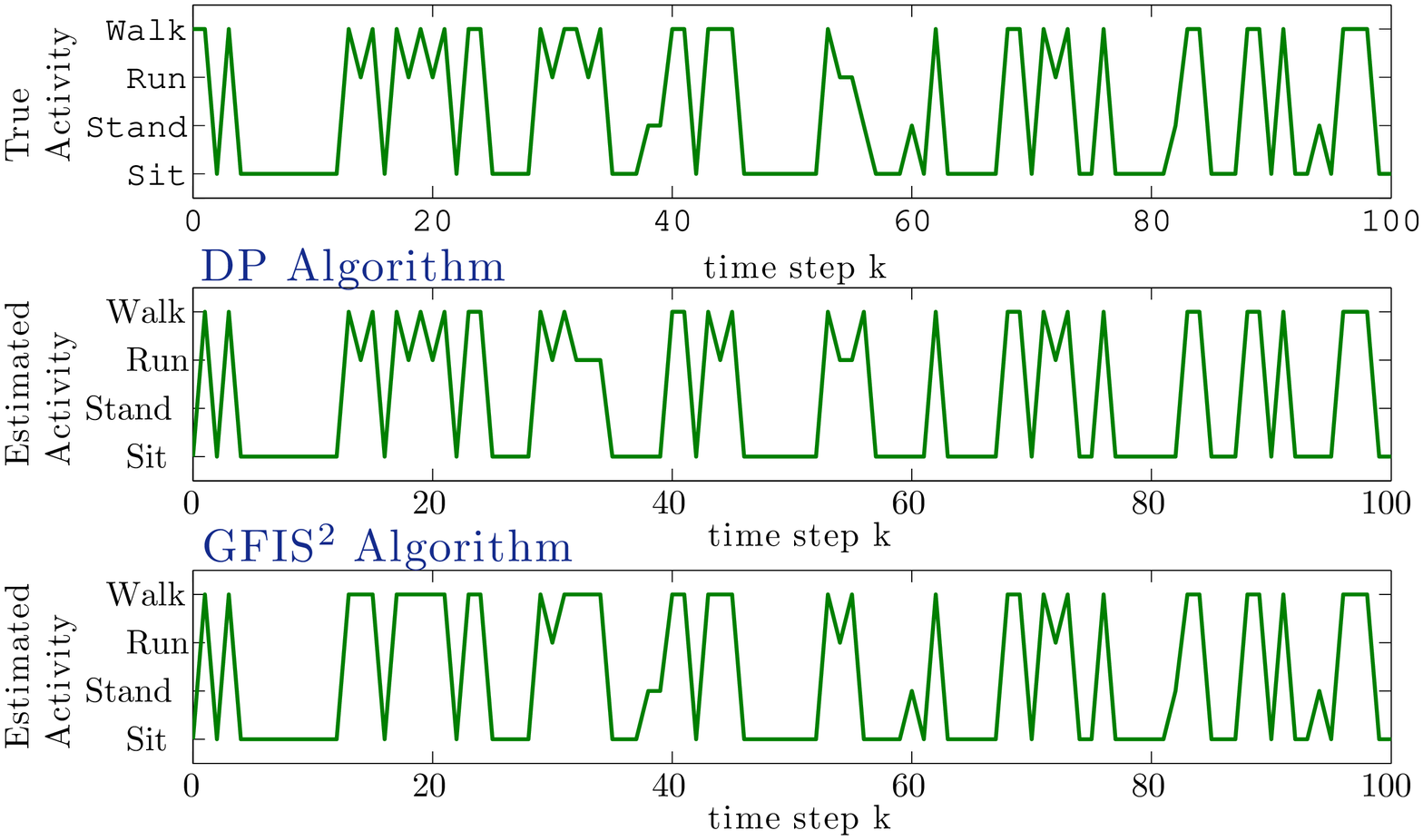}}
\caption{Tracking performance comparison. Top: true activity, middle: estimated activity (DP Algorithm), bottom: estimated activity (GFIS$^2$ Algorithm).}
\label{fig:tracking_performance}
\end{figure}

Fig.~\ref{fig:tracking_performance} depicts the true system state sequence and the tracking performance of DP and GFIS$^2$. We note that both algorithms track very well the individual's time--varying activity state despite the few number of samples used. We also observe that GFIS$^2$ usually confuses the \emph{Walk} and \emph{Run} states, which has to do with how close the associated signal models are in conjunction with the Markov chain transition probabilities values. However, its greedy nature benefits detecting the \emph{Stand} state versus DP, which optimizes the average cost and filters out states out with low stationary probability. Table \ref{tb:MSE_detection_accuracy} shows that the performance loss due to the adoption of GFIS$^2$ is small. Meanwhile, the associated reduction in complexity is significant, making the proposed algorithm attractive for controlled sensing applications. It is possible to achieve better MSE/detection accuracy by considering a Bayesian version \cite{VanTrees07} of the Fisher information measure and/or extensions to the dynamical case \cite{VanTrees07}. However, this may significantly increase the related complexity. 

\begin{table}[tb!]
\caption{MSE and Detection accuracy comparison between DP and GFIS$^2$ algorithms.}
\begin{center}
  \begin{tabular}{ | c || c | c |}
    \hline
    Sensing strategy & DP & GFIS$^2$\\ \hline
    MSE & $0.3791$ & $0.3848$ \\ \hline
    Detection accuracy & $87\%$ & $84\%$ \\ \hline
  \end{tabular}
\end{center}
\label{tb:MSE_detection_accuracy}
\end{table}

\begin{figure}[tb!]
\centering
{\includegraphics[width=\columnwidth]{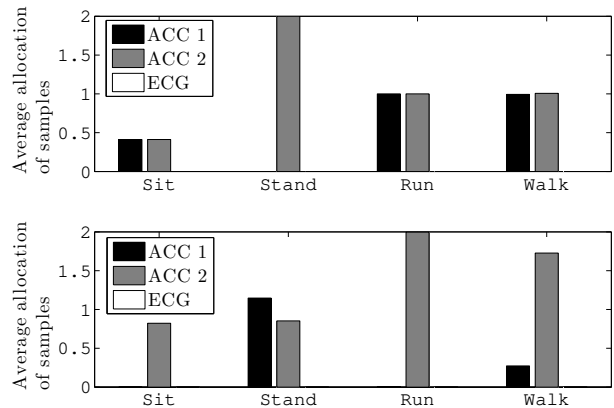}}
\caption{Average allocation of samples per sensor and state comparison. Top: DP Algorithm, bottom: GFIS$^2$ Algorithm.}
\label{fig:average_allocation_samples}
\end{figure}

Finally, Fig.~\ref{fig:average_allocation_samples} illustrates the average number of samples per sensor and per state selected by DP and GFIS$^2$. We notice that both algorithms request no samples from the ECG as expected, since according to Fig.~\ref{fig:4Hypothesis} the associated distributions are highly overlapping. On the other hand, both algorithms request a combination of samples from the two accelerometers, where the exact number depends on the underlying physical activity state and the adopted algorithm. An interesting observation is that GFIS$^2$ tends to select, on average, more samples from the second accelerometer in contrast to DP, which requests on average same number of samples from both ACCs. In contrast, for the $\emph{Stand}$ state, the situation is reversed, \emph{i.e.} GFIS$^2$ selects on average more samples from ACC 1 than ACC 2, while DP requests samples only from ACC 2.

\section{Conclusion}

In this paper, we considered the controlled sensing problem in the case of discrete--time, finite--state Markov chains with controlled Gaussian measurement vectors. Despite that the optimal solution is computationally intensive, it was possible to design a suboptimal, lower complexity algorithm based on the Fisher information measure. To this end, we generalized the Fisher information measure to account for multi--valued discrete parameters and control inputs. Numerical simulations using real data from a physical activity state tracking application indicated the near--optimality of the proposed algorithm. Future work will focus on determining theoretical performance guarantees for the proposed algorithm and considering sensor usage costs.

\bibliographystyle{IEEEtran}
\bibliography{refs}

\end{document}